\documentclass[aps,prd,showpacs,nofootinbib,10pt,superscriptaddress,twocolumn,amsmath,amssymb,floatfix,showkeys]{revtex4-1}
\usepackage{color,amsmath,amssymb,graphicx,latexsym,lineno}
\usepackage{threeparttable,multirow,txfonts,booktabs,bm}
\usepackage{accents,slashed,ulem,subfig}
\usepackage[para]{footmisc}
\definecolor{blue0}{rgb}{0,0,0.6}
\usepackage[colorlinks,linkcolor=blue0,anchorcolor=blue0,citecolor=blue0,urlcolor=blue0]{hyperref}
\usepackage{natbib}

\newlength{\dhatheight}
\newcommand{\doublehat}[1]{%
    \settoheight{\dhatheight}{\ensuremath{\hat{#1}}}%
    \addtolength{\dhatheight}{-0.15ex}%
    \hat{\vphantom{\rule{1pt}{\dhatheight}}%
    \smash{\hat{#1}}}}
\graphicspath{{figs/}}

\begin{document}

\title{Impact of Parameters in the Blazar Jet Magnetic Field Model on Axion-Like Particle Constraints}

\author{Lin-Qing Gao}
\affiliation{School of Nuclear Science and Technology, University of South China, Hengyang 421001, China}
\author{Xiao-Jun Bi}
\affiliation{Key Laboratory of Particle Astrophysics, Institute of High Energy Physics,
Chinese Academy of Sciences, Beijing 100049, China}
\affiliation{School of Physical Sciences, University of Chinese Academy of Sciences, Beijing, China}
\author{Jun Li}
\affiliation{Key Laboratory of Dark Matter and Space Astronomy, Purple Mountain Observatory, Chinese Academy of Sciences, 210033 Nanjing, Jiangsu, China}
\affiliation{School of Astronomy and Space Science, University of Science and Technology of China, 230026 Hefei, Anhui, China}
\author{Peng-Fei Yin}
\affiliation{Key Laboratory of Particle Astrophysics, Institute of High Energy Physics, Chinese Academy of Sciences, Beijing 100049, China}

\begin{abstract}

The interaction between axion-like particles (ALPs) and photons induces ALP-photon oscillations in astrophysical magnetic fields, leading to spectral distortions in the $\gamma$-ray spectrum of blazars. The primary uncertainty of this phenomenon may originate from the magnetic field within the jet of the blazar. While many studies have explored the effects of ALP-photon oscillations using typical values for jet magnetic field parameters, it is important to recognize  that these parameters can be constrained by multi-wavelength observations. In this study, we utilize the high energy $\gamma$-ray spectrum of Mrk 421 obtained from MAGIC and Fermi-LAT observations. By employing multi-wavelength fitting with a one-zone synchrotron self-Compton model, we derive the parameters characterizing the magnetic field model within the jet, and investigate their impacts on the ALP constraints.


\end{abstract}

\maketitle

\section{introduction}

Axion-like particles (ALPs) are pseudo-scalar particles predicted by a variety of extensions to the Standard Model \cite{Coriano:2006xh,Baer:2008yd}, Kaluza-Klein theories \cite{Turok:1995ai,Chang:1999si,Dienes:1999gw}, and especially superstring theories \cite{Svrcek_2006,Arvanitaki:2009fg,Cicoli:2012sz}. While ALPs exhibit similarities to axions, they do not necessarily offer a solution to the strong CP problem. Due to their potential as candidates for dark matter and their relevance to numerous crucial cosmological issues  \cite{Marsh:2015xka, Preskill:1982cy, Abbott:1982af, Dine:1982ah, Turner:1989vc, Sikivie:2006ni}, ALPs are actively being pursued in laboratory experiments, as well as through astrophysical and cosmological observations.

The interaction between the ALP and photons results in oscillations between them within external magnetic fields. Given the widespread presence of astrophysical magnetic fields along the trajectory from the source to Earth, the phenomenon of ALP-photon conversion has the potential to produce observable effects that can be explored through astrophysical experiments. Specifically, the expected consequences of ALP-photon oscillations include irregularities in the $\gamma$-ray spectrum of astrophysical sources and alterations in the transparency of very high-energy photons \cite{DeAngelis:2007dqd, Hooper:2007bq, Simet:2007sa, Mirizzi:2007hr, Belikov:2010ma, DeAngelis:2011id, Horns:2012kw, HESS:2013udx, Meyer:2013pny, Tavecchio:2014yoa, Meyer:2014epa, Meyer:2014gta, Fermi-LAT:2016nkz, Meyer:2016wrm, Berenji:2016jji, Galanti:2018upl, Galanti:2018myb, Zhang:2018wpc, Liang:2018mqm, Bi:2020ths, Guo:2020kiq, Li:2020pcn, Li:2021gxs, Cheng:2020bhr, Liang:2020roo, Xia:2018xbt, Gao:2023dvn, Gao:2023und, Li:2024zst, Li:2024ivs, MAGIC:2024arq}. Additionally, these oscillations may induce changes in the polarization state of photons \cite{Bassan:2010ya, Galanti:2022iwb, Galanti:2022yxn, Galanti:2022tow, Dessert:2022yqq, Yao:2022col, Gau:2023rct, POLARBEAR:2024vel}.

This study focuses on the spectral distortions in $\gamma$-ray emissions resulting from ALP-photon oscillations for blazars, which are prominent sources in the extragalactic $\gamma$-ray domain. Blazars are a subclass of active galactic nuclei characterized by a supermassive black hole encompassed by an accretion disk, emitting two jets orthogonal to the disk.  Initially, photons are emitted from a region near the central supermassive black hole. Subsequently, they may traverse through the magnetic fields of the jet, galaxy cluster, intergalactic space, and finally, the Milky Way.

The uncertainties arising from the magnetic field environments significantly impact the effects of ALP-photon oscillations. The primary uncertainty originates from the blazar jet magnetic field (BJMF) model. This magnetic field is typically characterized by two components: a poloidal and a toroidal component, with the latter exerting a predominant influence on regions situated further from the central region\cite{Pudritz:2012xj, Begelman:1984mw}. The toroidal component of the magnetic field is oriented perpendicular to the jet axis, and its strength can be described as $B(r) = B_0 (r/r_0)^{-1}$, where $B_0$ represents the magnetic field strength of the emitting region, and $r_0$ denotes the distance of the emitting blob from the central black hole \cite{Pudritz:2012xj, Begelman:1984mw, OSullivan:2009dsx}. These parameters characterize the magnetic field strength in the jet, and have a profound impact on the rate of ALP-photon conversion and the interpretation of observed data.  In this study, we focus on this specific BJMF model. Discussions on more complicated jet models and their implications can be found in Ref.~\cite{Davies:2020uxn,Davies:2022wvj}.

In many previous studies, the effects of ALP-photon oscillations have been explored by assuming certain typical values for the parameters of the BJMF model. However, it is crucial to recognize that the magnetic field within the emitting region plays an important role in shaping the multi-wavelength emissions from blazars. Within a simple one-zone synchrotron self-Compton (SSC) model, the spectral energy distribution (SED) is generated by a single emission zone. The typical SED consists of two distinct bump components \cite{Ghisellini:2017ico}: the  low-energy component is attributed to synchrotron radiation emitted by electrons, and the high-energy component originates from inverse Compton scattering of electrons with the synchrotron photons produced by the same electron population \cite{Celotti:2007rb,Ghisellini:2009fj}. Consequently, the parameters of the BJMF model can be constrained by fitting observational data obtained across various wavelengths.

In this study, we investigate the impact of the BJMF parameters on the ALP constraints through observations of Mrk 421, which is a BL Lac object located at a redshift of 0.031. Mrk 421 is recognized as one of the most luminous extragalactic $\gamma$-ray sources. We analyze the high-energy $\gamma$-ray spectrum of Mrk 421 using data from the Fermi-LAT and MAGIC experiments, collected during a particularly bright period in 2017, as detailed in Ref. \cite{MAGIC:2021zhk}.
We consider the uncertainties associated with the BJMF model obtained from multi-wavelength fitting. The ALP parameters are constrained through the application of the $\rm CL_s$ method \cite{Junk:1999kv,Read:2002hq_cls} as referenced in \cite{Gao:2023dvn, Gao:2023und}.

In Sec. \ref{sec:ALP-photon}, we briefly introduce the ALP-photon oscillation phenomenon in astrophysical environments. Sec. \ref{sec: SSC} focuses on the one-zone SSC model used to interpret the observations of Mrk 421. The fitting results of this model are presented. Sec. \ref{sec:method} explains the statistical method employed in our analysis, specifically the $\rm CL_s$ method used for setting constraints. The resulting constraints and the uncertainties are discussed in Sec. \ref{sec:results}. Finally, Sec. \ref{sec:conclusion} summarizes the conclusions drawn from our study.

\section{ALP-photon oscillation in astrophysical environments}\label{sec:ALP-photon}

The Lagrangian for the interaction between the ALP and photons can be expressed as
\begin{equation}
\mathcal{L}_{a\gamma} = -\frac{1}{4} g_{a\gamma} a F_{\mu \nu} \Tilde{F}^{\mu \nu} = g_{a\gamma} a \boldsymbol{E} \cdot \boldsymbol{B},
\end{equation}
where $a$ represents the ALP field, $F$ represents the electromagnetic field strength tensor, $\Tilde{F}$ is its dual, $g_{a\gamma}$ represents the coupling strength, and $\boldsymbol{E}$ and $\boldsymbol{B}$ represent the electric and magnetic fields, respectively. The conversion between ALPs and photons takes place in the presence of an external magnetic field.  We utilize a state vector $\boldsymbol{\Psi} = (A_1, A_2, a)^T$ to characterize the system propagating along the direction $z$, where $A_1$ and $A_2$ denote the photon polarization amplitudes.

The polarization state of the ALP-photon system is described by the density matrix $\rho \equiv \Psi \otimes \Psi^\dagger$. When traversing a uniform magnetic field, $\rho$ follows a von Neumann-like commutator equation 
\cite{DeAngelis:2007dqd,Mirizzi:2009aj} as
\begin{equation}\label{equ:von Neumannn-like}
    i\frac{d\rho}{d\boldsymbol{z}} = [\rho, \mathcal{M}_0], 
\end{equation}
where $\mathcal{M}_0$ represents the mixing matrix including the interactions between photons and external magnetic field. In the case of a transverse magnetic field $B_t$ along the direction $\boldsymbol{x_2}$, $\mathcal{M}_0$ is given by
\begin{equation}
    \mathcal{M}_0 = \begin{bmatrix}
\Delta_{\perp} & 0 & 0 \\
0 & \Delta_{\parallel} & \Delta_{a\gamma} \\
0 & \Delta_{a\gamma}  & \Delta_{a} 
\end{bmatrix},
\end{equation}
where the first two diagonal elements \(\Delta_{\perp}\) and \(\Delta_{\parallel}\) include various astrophysical and quantum effects without ALPs,
the third diagonal element $\Delta_{a}$ is given by \(\Delta_{a} = m_a^2/2E\), and the off-diagonal element \(\Delta_{a\gamma} = g_{a\gamma} B_t/2\) denotes the ALP-photon mixing effect. The plasma effects integrated into \(\Delta_{\perp}\) and \(\Delta_{\parallel}\) are delineated by \(\Delta_{pl} = -\omega_{pl}/2E\), where \(\omega_{pl} = (4\pi \alpha n_e/m_e)^{1/2} \) denotes the characteristic frequency of the plasma depending on the electron density $n_e$ of the surrounding environment.

For the initial state with $\rho(0)$, the solution to Eq.~\ref{equ:von Neumannn-like} can be expressed as \(\rho(\boldsymbol{z}) = \mathcal{T}(\boldsymbol{z}) \rho(0) \mathcal{T}(\boldsymbol{z})^\dagger\), where \(\mathcal{T}(\boldsymbol{z})\) represents the transfer matrix depending on the mixing matrix. 
The entire path of photons traversing multiple astrophysical magnetic fields can be segmented, with each small segment characterized by a constant magnetic field. The survival probability of these photons is given by \cite{Raffelt:1987im,Mirizzi:2007hr}:
\begin{equation}\label{equ:Pgaga}
P_{\gamma\gamma} = \mathrm{Tr}\left((\rho_{11}+\rho_{22})\mathcal{T}(\boldsymbol{z}) \rho(0) \mathcal{T}^{\dagger}(\boldsymbol{z})\right),
\end{equation}
where $\mathcal{T}(\boldsymbol{z}) = \prod \limits_i^n \mathcal{T}_i (\boldsymbol{z})$ represents the total transfer matrix, with $\mathcal{T}_i(\boldsymbol{z})$ derived from each segment along the photon's path. The initial photons are assumed to be unpolarized with $\rho(0) = \rm{diag}(1/2,1/2,0)$. $\rho_{11}$ and $\rho_{22}$ represent the density matrices corresponding to the photon polarization states, with $\rm{diag}(1,0,0)$ and $\rm{diag}(0,1,0)$, respectively.

High-energy photons emitted from the blazar traverse various astrophysical environments. Our analysis of Mrk 421 focuses primarily on the blazar jet, extragalactic space, and the Milky Way. 
By considering ALP-photon oscillations in the blazar jet and the Milky Way, as well as the effects of photon absorption by EBL photons in the extragalactic space, we can determine the survival probability $P_{\gamma\gamma}$ for high energy photons emitted from Mrk 421. In this investigation, we employ the package gammaALPs \cite{Meyer:2021pbp} to compute $P_{\gamma\gamma}$.

\section{Fitting to broadband multi-wavelength observations}\label{sec: SSC}

In order to obtain the parameters of the BJMF model, we conduct a broadband multi-wavelength model fitting using the observed SED data. 
The SED data of Mrk 421 collected in four distinct periods are provided by Ref. \cite{MAGIC:2021zhk}, encompassing observations ranging from radio wavelengths to very high-energy $\gamma$-rays obtained by instruments including MAGIC, FACT, Fermi-LAT, \textit{NuSTAR},  \textit{Swift}, GASP-WEBT, and OVRO. Notably, the observation recorded on MJD 57757 exhibits the highest very high-energy luminosity among the four periods. Consequently, our analysis is specifically focused on scrutinizing the SED data corresponding to MJD 57757.

The SED of the blazar exhibits two distinct components:  a low-energy component primarily attributed to synchrotron radiation resulting from interactions between relativistic electrons and the magnetic field, and a high-energy component that remains a topic of ongoing debate. In leptonic models, the high-energy component is attributed to either the inverse-Compton scattering of synchrotron photons emitted by the same electron population or thermal photons originating from surrounding structures. The former process is known as the SSC model~\cite{Celotti:2007rb, Ghisellini:2009fj}, while the latter is denoted as the external Compton model\cite{Dermer:1993cz, Blazejowski:2000ck}. In hadronic models, the high-energy component can be explained by proton synchrotron emission \cite{Aharonian:2000pv} or emissions from secondary particles generated in photohadronic ($p\gamma$) and hadronuclear ($pp$) interactions \cite{Mannheim:1993jg, Cerruti:2014iwa}, which can also lead to the production of high-energy neutrinos. 

The commonly employed one-zone leptonic model predicts that the multi-wavelength emission originates from a compact region filled with relativistic electrons, leading to correlations between light curves across different energy bands. However, in cases where correlations between light curves across different energy bands are absent, or the occurrence of orphan flares, the efficacy of one-zone leptonic models in providing a satisfactory explanation is challenging. In such scenarios, multi-zone leptonic models may offer a more comprehensive understanding \cite{MAGIC:2021zhk,Multi-wavelengthcollaborators:2024aaw,Liodakis:2019vmd,MAGIC:2019mfn,Krawczynski:2003fq}. Previous studies \cite{Finke:2008pe, Aleksic:2011ta, NuSTARTeam:2015jhz, MAGIC:2021zhk} have demonstrated the utility of the one-zone SSC model in interpreting multi-wavelength observations of Mrk 421 in most cases. In our study, we adopt this model as the intrinsic mechanism for interpreting the observed data. 

In the framework of the one-zone SSC model, the broadband SED is emitted from a single region, which is filled with relativistic electrons and contains a homogeneous magnetic field. This emitting region is assumed to have a spherical configuration with a radius denoted as \( R \), traversing relativistically with a bulk Lorentz factor \( \Gamma \). The strength of this magnetic field is characterized by \( B_0 \). The homogeneous electron population is described by a broken power-law distribution given by:
\begin{equation}
    N(\gamma) =
    \begin{cases}
    N_0 \gamma^{-\alpha_1}, & \gamma_{\text{min}} < \gamma < \gamma_{\text{br}} \\
    N_0 \gamma^{-\alpha_2} \gamma_{\text{br}}^{\alpha_2 - \alpha_1}, & \gamma_{\text{br}} < \gamma < \gamma_{\text{max}} \;\;,
    \end{cases}
\end{equation}
where \( N_0 \) is the normalization constant, and \( \gamma_{\text{min}} \), \( \gamma_{\text{br}} \), and \( \gamma_{\text{max}} \) correspond to the minimum, break, and maximum Lorentz factors of the electron population, respectively. The indices \(\alpha_1\) and \(\alpha_2\) represent the power-law indices below and above the break Lorentz factor, respectively. Based on this electron distribution, the energy density of the electrons can be calculated as \( U_e = m_e c^2 \int_{\gamma_{\text{min}}}^{\gamma_{\text{max}}} \gamma N(\gamma) \, d\gamma \).

The photon energy in the rest frame of the emitting region and the photon energy in the laboratory frame are related through the Doppler factor. The Doppler factor \(\delta_D\) is determined by the expression \(\delta_D = [\Gamma(1-\beta \cos \Theta)]^{-1}\), where \(\Gamma\) and \(\beta\) represent the Lorentz factor and velocity of the emitting region, respectively, and \(\Theta\) represents the angle between the jet and the line of sight. To simplify the model and reduce the number of free parameters, we assume \(\Theta=1/\Gamma\), leading to \(\delta_D\) = \(\Gamma\). In our analysis, we adopt a typical value of the Doppler factor, such as 25 \cite{Tavecchio:2009zb}, in the fitting process. Furthermore, the radius \(R\) of the emitting region is estimated to be \(10^{16}\) cm based on light variation timescales. Due to limited constraints from the available data, we set the minimum Lorentz factor \(\gamma_{\text{min}}\) to be 1000, which is consistent with that utilized in Ref. \cite{MAGIC:2021zhk}. With these parameters fixed, there remain 6 free parameters in the fitting process: \(B_0\), \(\gamma_{\text{br}}\), \(\gamma_{\text{max}}\), \(\alpha_1\), \(\alpha_2\), and \(N_0\).

\begin{figure}[ht]
  \centering
 \includegraphics[width=0.45\textwidth]{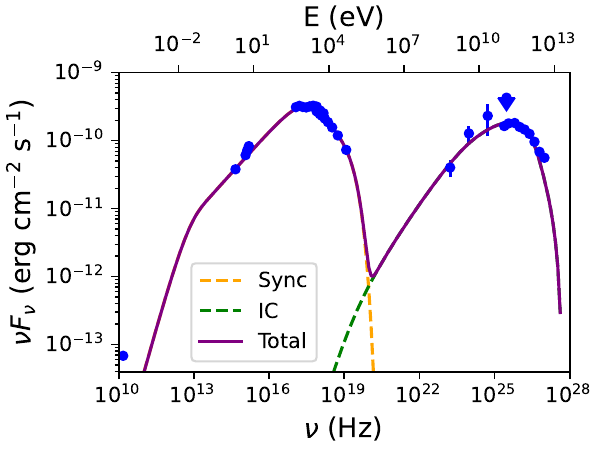}
  \caption{The SED of multi-wavelength observations on MJD 57757. The data points taken from \cite{MAGIC:2021zhk} are depicted by the blue points. The purple solid line represents the best-fit result of the one-zone SSC model with the parameter set I as detailed in Table \ref{tab:SSC_fit}. The dashed green and yellow lines correspond to the synchrotron and inverse Compton scattering spectra, respectively.}
  \label{fig:SSC}
\end{figure}

We utilize the publicly available code JetSeT \cite{2009A&A...501..879T,2011ApJ...739...66T,tramacere2020jetset} to conduct the SED fitting for multi-wavelength observations, including MAGIC, Fermi-LAT, \textit{NuSTAR}, \textit{Swift}, GASP-WEBT, and OVRO.
The best-fit spectra are illustrated in Fig. \ref{fig:SSC}, with the corresponding best-fit  parameters outlined in the first column of Tab. \ref{tab:SSC_fit}. The fitting procedure generally yields satisfactory results, with the potential exception of the radio band, where the discrepancy is attributed to the synchrotron self-absorbed effect. Notably, the derived parameters closely align with those reported in Ref. \cite{MAGIC:2021zhk}. In our analysis, the parameters $\delta_D$, $R$, and $\gamma_{\text{min}}$ are fixed in the fitting. Nevertheless, to accommodate for their uncertainties, we also present the fitting results using some distinct sets of $\delta_D$ and $R$, as listed in Tab. \ref{tab:SSC_fit}. Our results are consistent with the results derived from other observations \cite{MAGIC:2021zhk,ARGO-YBJ:2015qiq,LAT:2011mmt,MAGIC:2014czi}.

Note that the magnetic field strengths derived from hadronic models are significantly larger than those derived from leptonic models. When fitting the data using some specific hadronic models, the magnetic field strength inferred can range from several to tens of Gauss \cite{LAT:2011mmt,Zech:2017lma,Mastichiadis:2013aga}.  Consequently, the utilization of hadronic models could impose more stringent constraints on ALP.

\begin{table*}[t]
\centering
\begin{tabular}{ccccccccc}
\hline
\hline
Status & Parameters & I & II & III &  IV &  V\\
\hline
      & $\delta_D$ & 25 & 35 & 20 & 25 & 25 &  \\
fixed & R [$\times$ $10^{16}$ cm] & 1  & 1  & 1 & 3 & 0.5  \\
      & $\gamma_{\rm min}$ & 1000 & 1000 & 1000 & 1000 & 1000  \\ 
\hline
     & B [$\times$ $10^{-2} \rm{G}$] & 5.90 $\pm$ 0.34 & 3.40 $\pm$ 0.10 & 8.50 $\pm$ 0.54 & 1.97 $\pm$ 0.11  &  11.8 $\pm$ 0.73 \\
    & $N_0$ [$\times$ $10^{4}\rm {m}^{-3}$] & 1.40 $\pm$ 0.78  & 1.25 $\pm$ 0.68 & 1.60 $\pm$ 0.98 & 0.49 $\pm$ 0.24 & 3.05 $\pm$ 1.87   \\
free    & $\gamma_{\rm br}$ [$\times$ $10^{5}$] & 2.24 $\pm$ 0.26 & 2.43 $\pm$ 0.29 & 2.12 $\pm$ 0.25 & 3.87 $\pm$ 0.44 & 1.62 $\pm$ 0.18  \\
 & $\gamma_{\rm max}$ [$\times$ $10^{6}$] & 1.74 $\pm$ 0.82 & 1.73 $\pm$ 0.56 & 1.75 $\pm$ 0.82 & 2.77 $\pm$ 0.95 & 1.68 $\pm$ 1.61 \\
    & $\alpha_1$ & 2.18 $\pm$ 0.05 & 2.20 $\pm$ 0.05 & 2.17 $\pm$ 0.06 & 2.22 $\pm$ 0.05 & 2.16 $\pm$ 0.06 \\ 
     & $\alpha_2$ & 3.79 $\pm$ 0.17 & 3.75 $\pm$ 0.15 & 3.81 $\pm$ 0.14 & 3.77 $\pm$ 0.16 & 3.84 $\pm$ 0.16 \\
\hline
\hline
\end{tabular}
\caption{The parameters of the one-zone SSC models for observations on MJD 57757. The values of the first three parameters are fixed in the fitting process, while the subsequent parameters exhibit the best-fit values accompanied by their respective uncertainties.}
\label{tab:SSC_fit}
\end{table*}

\section{method}\label{sec:method}

In this section, we introduce the SED fitting method and describe the statistical approach used to constrain ALP parameters, employing the $\rm CL_s$ method. We consider the multi-wavelength observational SED of Mrk 421 on MJD 57757 provided in Ref. \cite{MAGIC:2021zhk}. Given the enhanced efficiency of the ALP-photon oscillation effect in the $\gamma$-ray band, our investigation of ALP is specifically focused on the Fermi-LAT and MAGIC observations.

In theory, the intrinsic spectrum of the blazar can be directly predicted within the framework of a detailed radiation model. Nevertheless, due to the potential effects of ALPs and uncertainties related to astrophysical phenomena,  we characterize the intrinsic spectrum of the blazar in the $\gamma$-ray band using a log-parabolic function, denoted by $\Phi_{\text{int}} = F_0(E/E_0)^{-\alpha - \beta\mathrm{log} (E/E_0)}$, where
$\bm{\theta} \equiv\{\alpha, \beta, F_0\}$ 
are the spectral parameters, with $E_0$ fixed at 200 GeV. We take $\alpha=2.0$ as a default choice, that well reproduces the SED up to $\mathcal{O}(10^2)$GeV. Subsequently, by incorporating the ALP-photon oscillation effect and EBL absorption, we determine the photon survival probability $P_{\gamma\gamma}$. The resultant spectrum is then derived as $\Phi = \Phi_{\text{int}} \times P_{\gamma\gamma}$.

The optimal fit spectrum is determined by minimizing the $\chi^2$ function defined as
\begin{equation}\label{eq: chi2 def}
    \chi^2 = \sum\limits _{i=1}^{N} \frac{ (\Phi_{{\rm pre},i} - \Phi_{{\rm obs}, i} )^2 }{\delta \Phi_{{\rm obs}, i}^2}
\end{equation}
where $\Phi_{{\rm pre},i}$, $\Phi_{{\rm obs},i}$, and $\delta\Phi_{{\rm obs},i}$ denote the predicted value, observed value, and experimental uncertainty of the photon flux the $i$-th energy bin, respectively. Considering the energy resolution of the detector, the impact of ALP-photon oscillations is smoothed out in the observations. Therefore, to account for this effect in the predicted spectra, we average the expected values within each energy bin.

We define the test statistic (TS) as
\begin{equation}\label{equ:TS_CLs}
    {\rm TS}(m_a, g_{a\gamma}) = \chi^2_{\rm ALP} (\doublehat{\bm{\theta}}; m_a, g_{a\gamma})-\chi^2_{\rm Null}(\hat{\bm{\theta}}),
\end{equation}
where $\chi^2_{\rm Null}$ represents the best-fit $\chi^2$ under the null hypothesis without the ALP effect, and $\chi^2_{\rm ALP}$ denotes the best-fit $\chi^2$ value with the ALP effect at the specified parameters ($m_a$,$g_{a\gamma}$) under the alternative hypothesis. The parameters $\doublehat{\bm{\theta}}$ and $\hat{\bm{\theta}}$ correspond to the best-fit parameters of the intrinsic spectrum under the null and alternative hypotheses, respectively. 

In this study, we employ the ${\rm CL_s}$ method \cite{Junk:1999kv,Read:2002hq_cls,Lista:2016chp} to constrain the ALP parameters. The comprehensive elucidation of this approach can be found in the studies of  \cite{Gao:2023dvn,Gao:2023und,Li:2024ivs}; hence, we provide a concise overview here.
We scan the parameter points within the ($m_a, g_{a\gamma}$) plane, and evaluate their consistency with observational data. Here we take a specific parameter point ($m_{a,i}, g_{a\gamma,i}$) as an example to elucidate the method. Initially, utilizing Eq. \ref{eq: chi2 def}, we calculate the best-fit predicted flux values in each energy bin under both the null and  alternative hypotheses for the chosen parameter point.  Subsequently, we generate two mock data sets, $\rm \{d\}_{s+b}$ and $\rm \{d\}_{b}$, each containing 5000 samples, using the Gaussian sampling method. In the generation, the mean value and standard deviation are set to be the best-fit predicted flux value and observational uncertainty in each energy bin. Following this, we construct two distributions of the TS values, $\rm \{TS\}_{b}$ and $\rm \{TS\}_{s+b}$, using Eq. \ref{equ:TS_CLs}. Finally, by comparing these TS distributions with the TS value ${\rm TS_{obs}}$ obtained from the actual observational data, we determine the ${\rm CL_s}$ value. This value is defined as $\rm CL_s=CL_{s+b}/CL_{b}$, where $\rm CL_{s+b}$ and $\rm CL_b$ represent the probabilities of finding a TS value larger than ${\rm TS_{obs}}$ according to the distributions $\rm \{TS\}_{s+b}$ and $\rm \{TS\}_{b}$, respectively. If the ${\rm CL_s}$ value falls below 0.05, the selected parameter point is deemed excluded at a 95\% confidence level (C.L.).

\section{results}\label{sec:results}

\begin{figure}[ht]
  \centering
 \includegraphics[width=0.45\textwidth]{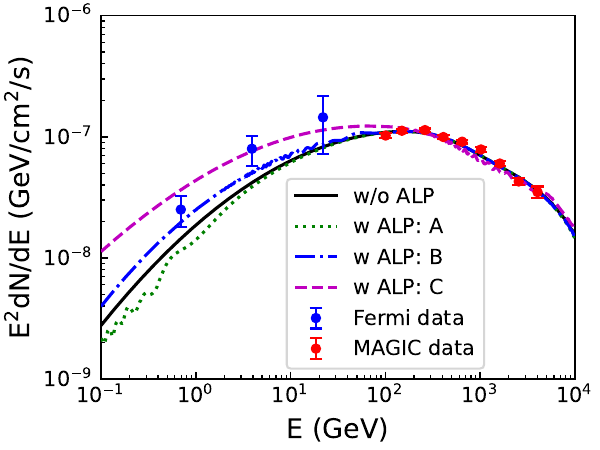}
  \caption{The high energy photon spectra under the null and alternative hypotheses. The ALP parameter points A, B, and C represent $(m_a, g_{a\gamma})= ({10^{-9} ~ \rm eV, \;  8\times10^{-11} ~ \rm GeV^{-1}})$, $(10^{-8} ~ \rm eV, $$\;  2.5\times10^{-11} ~\rm GeV^{-1})$, and $({10^{-7} ~\rm eV, \;  2.5\times10^{-11} ~\rm GeV^{-1}})$, respectively. The SED data observed by Fermi-LAT and MAGIC on MJD 57757 are represented by the red and blue points, respectively.}
  \label{fig:SED}
\end{figure}

In this section, we present the constraints on ALP obtained from the observations by MAGIC and Fermi-LAT on MJD 57757 for Mrk 421. 
The solid black line in Fig. \ref{fig:SED} shows the best-fit spectrum without considering ALPs. Incorporating the ALP effects, we compute the photon survival probabilities and derive the high energy photon spectra. The results for three benchmark points are depicted in Fig. \ref{fig:SED}. The default values of the BJMF parameters are based on the parameter set I in Tab.~\ref{tab:SSC_fit}.
The Doppler factor of the emitting region $\delta_D$ is set to 25, and the typical radius of the emitting region radius $R$ is taken to be $10^{16}~{\rm cm}$. Utilizing the assumption $r_0 = R/\Theta$, we obtain the distance $r_0 = 2.5 \times 10^{17}~{\rm cm}$. The one-zone SSC model yields a magnetic field strength of $B_0 = 0.059~{\rm G}$. As the electron density $n_0$ has minor impacts on the ALP effects, it is fixed at $n_0 = 3000~{\rm cm}^{-3}$ for all the calculations in this work. 

\begin{figure}[ht]
  \centering
 \includegraphics[width=0.45\textwidth]{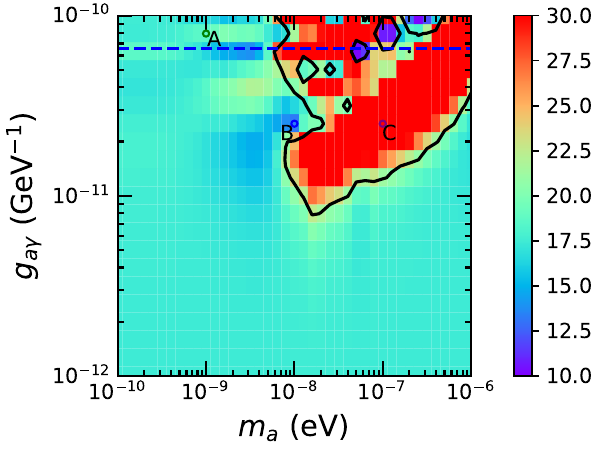}
  \caption{The heat map  illustrating the  $\chi^2$ values in the $m_a-g_{a\gamma}$ plane. The solid black lines represent the 95\% C.L. constraints determined through the $\rm CL_s$ method. The parameters of the BJMF model are based on the parameter set I as outlined in Tab.~\ref{tab:SSC_fit}. The dashed line represents the constraints from the CAST experiment~\cite{CAST:2017uph}.  Additionally, the parameter points A, B, and C, corresponding to the spectra displayed in Fig.~\ref{fig:SED}, are also shown.}
  \label{fig:best-fit}
\end{figure}

In order to explore the parameter space of ALP, we conduct a scan in the region of $m_a \in 
[10^{-10},10^{-6}]~\mathrm{eV}$ and $g_{a\gamma} \in [10^{-12},10^{-10}]~\mathrm{GeV}^{-1}$. Employing the $\rm CL_s$ method, we establish constraints on the ALP parameters at a 95\% C.L.. The corresponding constraints are illustrated by black solid lines in Fig. \ref{fig:best-fit}, with a color heat map indicating the $\chi^2$ value for each $(m_a, g_{a\gamma})$  under the alternative hypothesis.

\begin{figure}[htbp]
  \centering
  \begin{minipage}[b]{0.4\textwidth} 
    \centering
    \includegraphics[width=\textwidth]{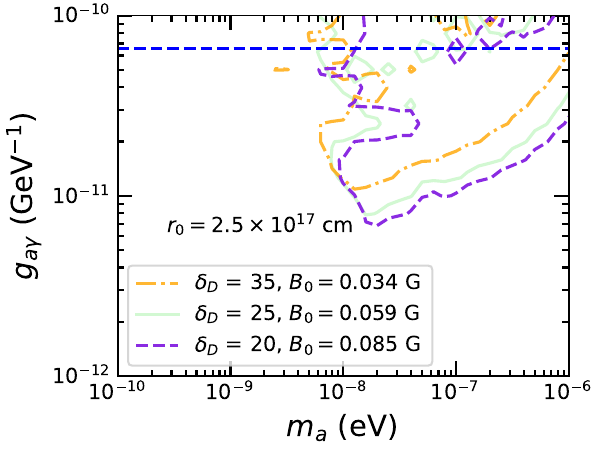}
    \caption*{(a)}
    \label{fig: B0 r0 err (a)}
  \end{minipage}
  \begin{minipage}[b]{0.4\textwidth} 
    \centering
    \includegraphics[width=\textwidth]{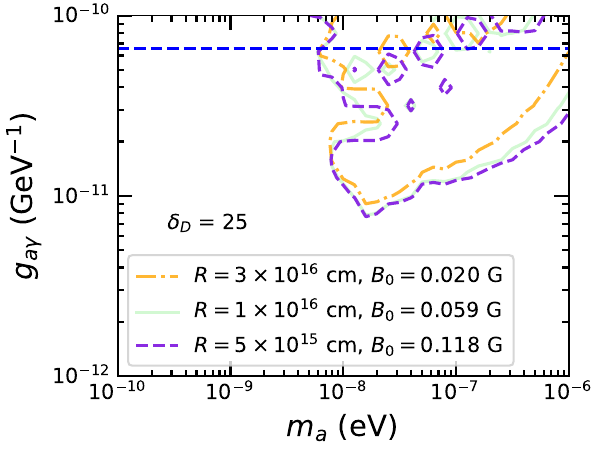}
    \caption*{(b)}
  \end{minipage} 
  \begin{minipage}[b]{0.4\textwidth} 
    \centering
    \includegraphics[width=\textwidth]{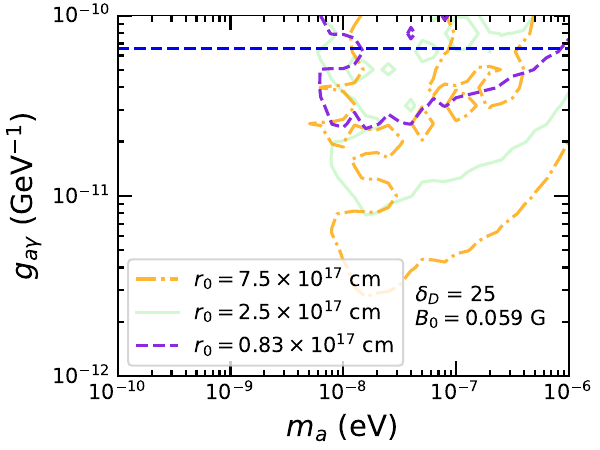}
    \caption*{(c)}
  \end{minipage}
  \caption{The constraints on the ALP parameters for different sets of parameters in the BJMF model as detailed in Tab.~\ref{tab:SSC_fit}: (a) constraints for the parameter sets I, III, and IV; (b) constraints for the parameter sets I, IV, and V; (c) constraints for the parameter set I with $r_0=7.5 \times 10^{17}$, $2.5\times 10^{17}$, and $0.83\times 10^{17}$ cm.}
  \label{fig:err}
\end{figure}

In the context of fitting multi-wavelength SED with one-zone SSC model, it is customary to maintain the parameters $\delta_D$ and $R$ at fixed values. In order to evaluate the impact of modifying these parameters on the fitting results and subsequent ALP constraints, we conduct an exploration of different values for $\delta_D$ and $R$. Specifically, we consider two distinct values for the Doppler factor $\delta_D = 20$ and 35.
The size of the emitting region is constrained by the light crossing time, given by $R \le \delta_D t_{\text{var}} c/(1+z)$, where $t_{\text{var}}$ is the variability time scale in the laboratory frame. $t_{\text{var}}$ is determined to be $t_{\text{var}}\sim 4-11$ hours in Ref.~\cite{MAGIC:2021zhk}. Accordingly, we consider two specific values for $R$: $5\times10^{15} ~{\rm cm}$ and $3\times 10^{16} ~{\rm cm}$.
Under the assumption $r_0 = R/\Theta \sim R \delta_D$, we obtain the corresponding values for $r_0$ to be $1.25\times10^{16}~{\rm cm}$ and $7.5\times 10^{17}~{\rm cm}$, respectively. The fitting results using the one-zone SSC model are summarized in Tab. \ref{tab:SSC_fit}. Notably, alterations in $\delta_D$ and $R$ affect the multi-wavelength fitting, thereby influencing the value of $B_0$. The results depicted in the top and middle panels of Fig. \ref{fig:err} correspond to the constraints obtained with varying $\delta$ and $R$, respectively. Within the range of variations explored, the figures demonstrate that both the Doppler factor and $R$ have a minor impact on the constraints. From Tab.\ref{tab:SSC_fit}, we note that when $R$ increases or decreases by a factor of $\mathcal{O}(1)$, $B_0$ undergoes a corresponding change in the opposite direction, resulting in similar constraint as shown in Fig. \ref{fig:err} (b).

The relationship $r_0 = R/\Theta$ should be considered as an initial estimate rather than a definitive assertion. To further investigate the impact of altering $r_0$ on the constraints, we maintain $R$ as a constant while varying $r_0$. Constraints are established for distinct $r_0$ values $0.83\times10^{16} ~{\rm cm}$ and $7.5\times 10^{17} ~{\rm cm}$, with the corresponding results depicted in the lower panel of Fig. \ref{fig:err}.  The constraints in Fig. \ref{fig:err} (c) demonstrate that higher values of $r_0$ result in notably more stringent constraints. Thus the most substantial uncertainty impacting the constraints arises from the uncertainty in $r_0$. If 
$r_0$ can be determined with high precision, the constraints will attain enhanced reliability. 

The direct impact of $\gamma_{\rm min}$ on the ALP constraints can be omitted, as fluctuations in the electron density $n_0$ within the mixing matrix can be neglected.
Nevertheless, this parameter may potentially influence the value of $B_0$ through the multi-wavelength fitting process. Note that the parameters $B_0$ and $r_0$ appear as a product in the expression of the BJMF, suggesting that varying the central value of $B_0$ mimics the effects of varying $r_0$.

\begin{figure}[h]
  \centering
\includegraphics[width=0.45\textwidth]{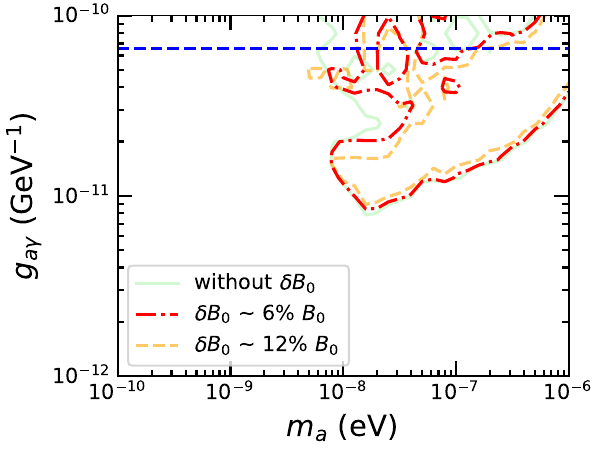}
  \caption{The constraints extended to include the uncertainties associated with $B_0$.
  The red dash-dotted line represents the constraints obtained considering the uncertainty of $\delta  B_0 \sim 6\% ~ B_0$ as derived from the one-zone SSC model fitting. The yellow dashed line represents the constraints derived under the assumption of a doubled uncertainty in $B_0$ as $\delta B_0 \sim 12\% ~ B_0$. For comparison, the constraints unaffected by the uncertainties in $B_0$ are also represented by the green solid line.}
  \label{fig:best-fit-B0err}
\end{figure}

The uncertainty of $B_0$ derived from the multi-wavelength fitting is detailed in Tab. \ref{tab:SSC_fit}. It is necessary to evaluate the influence of this uncertainty on the constraints. We adapt the $\chi^2$ function to include an additional term for varying magnetic filed strength:
\begin{equation}\label{eqB}
\chi^2 = \sum\limits_{i=1}^{N} \frac{(\Phi_{{\rm pre},i} - \Phi_{{\rm obs}, i})^2}{\delta \Phi_{{\rm obs}, i}^2} + \frac{(B'_0 - B_0)^2}{\delta B_{0}^2},
\end{equation}
where $B_0$ and $\delta B_{0}$ denote the central value of the magnetic field and its uncertainty derived from the multi-wavelength fitting, respectively. In this context, an additional free parameter $B'_0$ is introduced in the fitting process alongside the spectral parameters. Consequently, the survival probability $P_{\gamma\gamma}$ should be adjusted for each $B'_0$ explored during the fitting process. The subsequent steps follow the same method as in the previous analysis. The constraints resulting from this analysis are illustrated by the solid red dash-dotted lines in Fig. \ref{fig:best-fit-B0err}. Upon comparison with Fig. \ref{fig:best-fit}, it becomes apparent that the influence of the magnetic field uncertainty is relatively minor, attributed to the modest magnitude of $\delta B_{0}=3.4\times 10^{-3}\rm ~G \sim 6\% ~ B_0$. To further explore its sensitivity, we have increased $\delta B_{0}$ to $\delta B_{0}\sim 12\% ~B_0$ and evaluated its impact on the constraints, as shown by the dashed orange lines in Fig. \ref{fig:best-fit-B0err}. The constraints demonstrate a slight weakening even with a doubling of the uncertainty.

\begin{figure}[h]
  \centering
\includegraphics[width=0.45\textwidth]{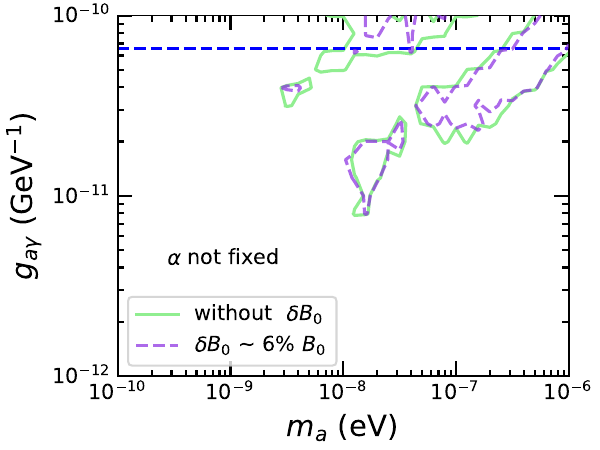}
  \caption{The constraints obtained with a freely varying parameter $\alpha$ of the intrinsic log-parabolic spectrum. The green and purple solid lines depict the constraints without and with the uncertainty of $B_0$ derived from the multi-wavelength fitting, respectively.}
  \label{fig:best-fit-B0err_free_alpha}
\end{figure}

In the preceding analysis, we have set the parameter $\alpha$ of the intrinsic log-parabolic spectrum to be 2.0, consistent with the multi-wavelength fitting in the high-energy band. For comparison, we also present the constraints obtained with a free $\alpha$, depicted by the green solid lines in Fig. \ref{fig:best-fit-B0err_free_alpha}. These constraints are weaker than those obtained with the fixed parameter $\alpha = 2.0$, as an additional nuisance parameter is introduced in the fitting process. Furthermore, as shown in Fig. \ref{fig:best-fit-B0err_free_alpha}, with the consideration of the uncertainty of $B_0$, the constraints do not vary significantly.

\begin{figure}[h]
  \centering
 \includegraphics[width=0.45\textwidth]{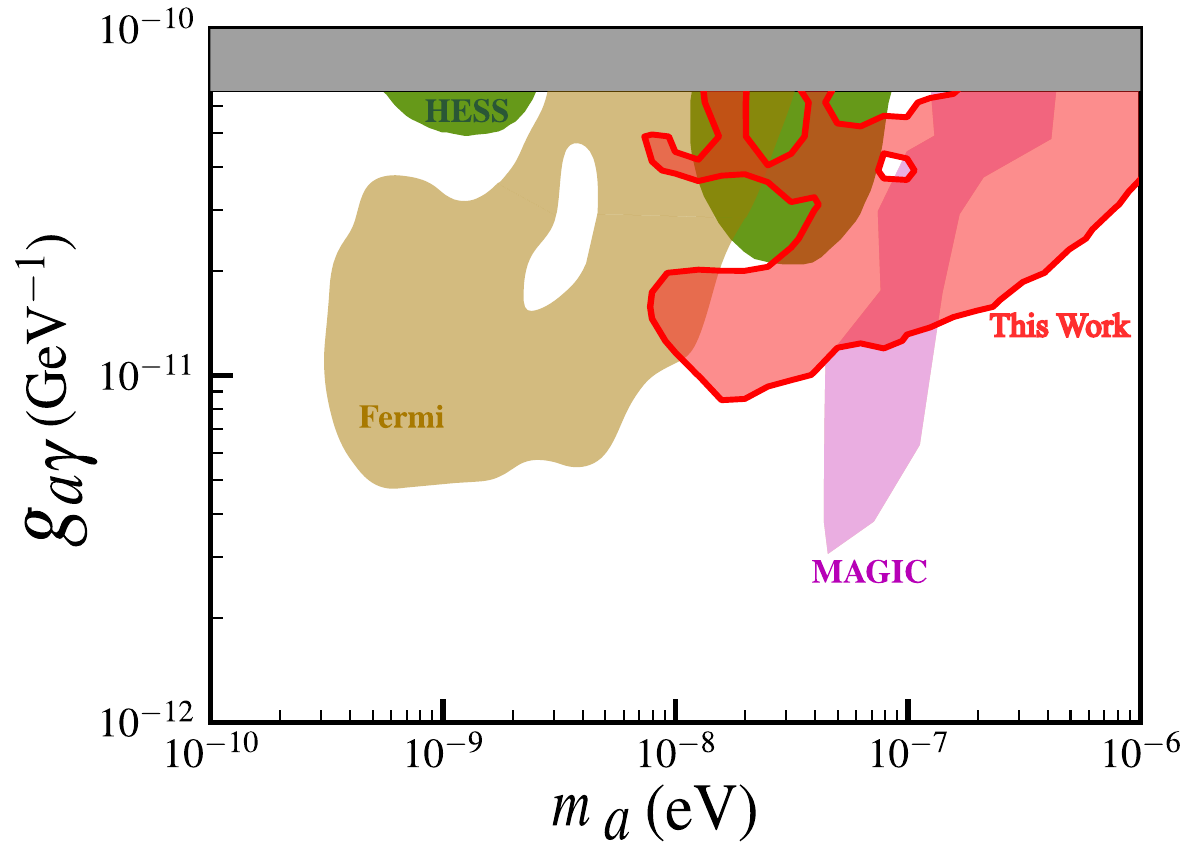}
  \caption{The red contours represent the 95\% C.L. constraints on ALP parameters including the uncertainty of $B_0$. Additionally, we compare these constraints with those from other observations, such as the CAST experiment \cite{CAST:2017uph}, the H.E.S.S. observation of PKS 2155-304 \cite{HESS:2013udx}, the Fermi-LAT observation of NGC 1275 \cite{Fermi-LAT:2016nkz}, and the MAGIC observation of the Perseus cluster \cite{MAGIC:2024arq}. Further constraints on ALP can be found in Ref. \cite{AxionLimits}.}
  \label{fig:comp}
\end{figure}

In Fig. \ref{fig:comp}, a comparison is presented between the constraints derived in this investigation and those from alternative observations, such as the CAST experiment \cite{CAST:2017uph}, the H.E.S.S. observation of PKS 2155-304 \cite{HESS:2013udx}, the Fermi-LAT observation of NGC 1275 \cite{Fermi-LAT:2016nkz}, and the MAGIC observation of the Perseus cluster \cite{MAGIC:2024arq}. The findings suggest that the constraints acquired in this study complement those from previous experimental research, particularly within the ALP mass range $10^{-8}-10^{-6}$ eV. 

\section{conclusions}\label{sec:conclusion}

In this study, we investigate the impact of parameters of the BJMF model on the ALP constraints. The high-energy photons emitted from blazars traverse diverse magnetic field environments before arriving at Earth, thereby introducing inherent uncertainties that affect ALP-photon oscillations. The primary uncertainty arises from the BJMF model. Previous studies predominantly explored the effects of ALP-photon oscillations by adopting specific values for the parameters characterizing the BJMF. However, it is important to recognize that these parameters are intricately interconnected and can be effectively constrained by multi-wavelength observations.

Employing a one-zone SSC model to fit multi-wavelength observations, we derive the magnetic field strength $B_0$ within the emitting region. Throughout the process of multi-wavelength fitting, we maintain the Doppler factor $\delta_D$ and the emitting region radius $R$ as constants. Our investigation further extends to the examination of varying $\delta_D$ and $R$ to evaluate their impacts on constraints, revealing their minor influences. We explore the influence of the distance from the emitting region to the central black hole $r_0$, and find that variations in $r_0$ lead to a more pronounced effect on constraints compared to other parameters under consideration. Furthermore, we incorporate the uncertainty associated with $B_0$ derived from the multi-wavelength fitting in the definition of $\chi^2$, thus evaluating its impact on the constraints. We find that the impact of this uncertainty on the constraints is insubstantial, owing to its relatively modest magnitude. 

Our analysis for ALPs focuses on the high-energy $\gamma$-ray observations of Mrk 421 on MJD 57757 conducted by Fermi-LAT and MAGIC. Utilizing the $\rm CL_s$ method, we establish constraints on the ALP parameters. The constraints derived in this study are complementary with those from previous experimental research within the ALP mass range $10^{-8}-10^{-6}$ eV.

\acknowledgments
This work is supported by the National Natural Science Foundation of China under grant No. 12175248.

\bibliographystyle{apsrev}
\bibliography{refs}

\end{document}